# Context-Aware Short-Term Interest First Model for Session-Based Recommendation

Haomei Duan and Jinghua Zhu

School of Computer Science and Technology,
Heilongjiang University, Harbin, China

## ABSTRACT

*In the case that user profiles are not available, the recommendation based on anonymous session is particularly important, which aims to predict the items that the user may click at the next moment based on the user's access sequence over a while. In recent years, with the development of recurrent neural network, attention mechanism, and graph neural network, the performance of session-based recommendation has been greatly improved. However, the previous methods did not comprehensively consider the context dependencies and short-term interest first of the session. Therefore, we propose a context-aware short-term interest first model (CASIF).The aim of this paper is improve the accuracy of recommendations by combining context and short-term interest. In CASIF, we dynamically construct a graph structure for session sequences and capture rich context dependencies via graph neural network (GNN), latent feature vectors are captured as inputs of the next step. Then we build the short-term interest first module, which can to capture the user's general interest from the session in the context of long-term memory, at the same time get the user's current interest from the item of the last click. In the end, the short-term and long-term interest are combined as the final interest and multiplied by the candidate vector to obtain the recommendation probability. Finally, a large number of experiments on two real-world datasets demonstrate the effectiveness of our proposed method.*

## 1. INTRODUCTION

In this era of the explosive growth of data, it is difficult for people to select the items they are interested in from a large number of items, and the recommendation system can recommend items that may be of interest to users. The recommendation system has been applied in many fields, such as e-commerce, music, and social media. It can be seen that the importance of the recommendation system. Most of the existing recommendation systems are based on the user's historical interactive information and make full use of the user's personal information. However, in some cases, users anonymously visit the site, thus the user's personal information is not available, besides, the user's historical interaction sequence is also very few. If the traditional method is used, it obviously is impossible to accurately recommend items for users. To solve this problem, session-based recommendation [1] is proposed to predict the next item that a user may click based on the sequence of the user's previous behaviors in the current session.

Due to the great practical value of session-based recommendation, it has been paid more and more attention, then various related recommendation algorithms are proposed. Markov Chain (MC) is a classic case, which assumes that the next action is based on the previous ones [2]. The main problem with Markov methods, however, is that they assume too strongly the independence of more than two consecutive actions, so a great number of important sequential information cannot be well exploited for sessions with more than two actions. The proposal of recurrent neural network (RNN) has obtained promising results in the session-based recommendation system, which has been proved to be effective in capturing users' preference from a sequence of historical actions [3, 4, 5]. NARM [6] is designed to capture the user's sequential pattern and main purpose simultaneously by employing a global and local RNN. However, RNN-based methods hold that there is a strict sequential relationship between two adjacent items in a session, which restricts the extraction of the characteristics of session dynamic changes.

After the successful application of Transformer [7] in Natural Language Processing (NLP), many attention-based models have been designed, which has been shown that comparable performance with RNNs in many sequence processing tasks. For the session-based recommendation tasks, a short-term attention/memory priority (STAMP) model [8] has been proposed to learn users' current interest and general interest in a session. However, it only takes the future mean value of all items in the session as the context, without taking the dynamic variations and local dependencies of the sequence into account. A position-aware context attention (PACA) model[9] for session-based recommendation has was proposed in 2019, which takes into account both the context information and the position information of items. However, this method only trains an additional implicit vector for each item and has little performance improvement for session-based recommendation tasks.

To tackle the above problems, we propose a context-aware short-term interest first model (CASIF) for session-based reccommendation, which takes into account both the locally dependent context information, long-term interest, and short-term interest on the whole. Due to graph neural network (GNN) [10] is capable of providing rich local contextual information by encoding edge or node attribute features, we dynamically construct a graph structure for session sequences and capture context dependencies via GNN. Based on the session graph, the proposed CASIF can capture transitions of neighbor items and generate the latent factor vectors for all nodes included in the graph. Then we build the short-term interest first module, which can capture the user's general interest from the session in the context of long-term memory, at the same time get the user's current interest from the item of the last click. The implicit vectors representing general interest and short-term interest pass through the multilayer perceptron respectively. In the end, the short-term and long-term interest are combined as the final interest and multiplied by the candidate vector to obtain the recommendation probability.

The main contributions of this work are summarized as follows.

- To represent the session characteristics more accurately, we present a novel context-aware short-term interest first model(CASIF) for session-based recommendation. CASIF fully utilizes the complementarity between short-term interest first attention and graph neural network to enhance the recommendation performance.
- The module based on graph neural network is used to model local graph-structured dependencies of separated session sequences, while short-term interest first module is designed to capture contextualized global representations.
- We conduct extensive experiments on two baseline datasets. Our experimental results show the effectiveness and superiority of CASIF, comparing with the state-of-the-art methods via comprehensive analysis.

The rest of the article is structured as follows. We will state relevant work in Section 2. Detailed our proposed context-aware short-term interest first model in Section 3, Section 4 presents our detailed experimental results and analysis, and finally, we conclude this paper in Section 5.

## 2. RELATED WORK

Session-based recommendation tasks are performed based on the user's anonymous historical behavior sequence and implicit feedback data, such as clicks, browsing, purchasing, etc., rather than rating or comment data. The primary aim is to predict the next behavior based on a sequence of the historical sequence of the session. The related works of session-based recommendation are summarized as follows.

### 2.1. Conventional methods

The algorithms based on decision rules [11, 12] or to train the prediction model with shallow features [13] are the simplest methods. However, their recommendation performance is poor. Matrix factorization [14, 15] is a general method of recommending systems, which basic aim is to decompose the user-item rating matrix into two low-rank matrices, each of which represents the latent factors of the user or item. However, because session-based recommendation is a sequential recommendation problem without a user profile, these traditional underlying factor models may not well suited for session-based recommendations. Many sequential recommendation methods based on the Markov chain (MC) [16] model predict the next item based on the previous one through computing transition probabilities between two consecutive items. FPMC [2] models the sequence behavior between every two adjacent clicks and provides a more accurate prediction for each sequence by factoring the user's personalized probability transfer matrix. However, the main disadvantage of Markov chain-based models is that the assumption of independence is too strong, which limits the accuracy of recommendations.

### 2.2. Deep learning methods

The successful application of deep learning in other fields has led many people to introduce related methods into session-based recommendation tasks. The most typical example is that Hidasi et al. [3] introduction of recurrent neural network (RNN) into session-based recommendation for the first time, which is called GRU4REC, and achieve significant progress over conventional methods. Because of their excellent performance, many followers began to try this method. Such as, Tan et al. [4] further propose two techniques to improve the performance of session-based recommendation. Although these methods have improved the performance of session-based recommendation, they are all restricted by the constraints of RNNs that both the offline training and the online prediction process are time-consuming, due to its recursive nature which is hard to be parallelized [18]. Recently, attention mechanisms have shown significant improvement in many machine learning tasks, such as machine translation [7], knowledge graph [17]. For the session-based recommendation task, more and more methods are proposed to utilize the attention mechanism to improve performance. Li et al. [6] propose a hybrid RNN-based encoder with an attention layer, which is called neural attentive recommendation machine (NARM), employs the attention mechanism on RNN to capture users' features of sequential behavior and main purposes. Then, a short-term attention priority model (STAMP) [8] using simple MLP networks and an attentive net, is proposed to efficiently capture both users' general interest and current interest. Xu et al. [31] proposed Graph Contextualized Self Attention Network for Session-based Recommendation, which is a combination of GNN and attention mechanisms, and further improves the accuracy of the recommendation.

## 2.3. Neural network on graphs

In recent years, neural networks have been used to generate representations of graphically structured data, such as social networks and knowledge bases [19, 20]. Besides, classic neural networks CNN and RNN are also deployed on graph structure data [21]. Previously, in the form of recurrent neural networks, graph neural networks (CNN) [22] proposed to operate on digraph. Gated GNN [23] is a modification of GNN that USES gated recursive units and USES time backpropagation (BPTT) to calculate gradients. In recent years, GNN has been widely applied to different tasks, such as script event prediction [24], scene recognition [25], image classification [26]. Wu et al. [27] have applied graph neural network (GNN), to extract item embedding from a session graph. Furthermore, items' embeddings are inputted into an attentive network to generalize the final representation for item prediction. Wang et al. [30] proposed a novel Multirelational Graph Neural Network model for Session-based target behavior Prediction, which obtained excellent results by modeling multi-relational item graph.

In a word, these deep neural networks and graph neural network models have a strong ability to extract comprehensive features of depth, which greatly improves the performance of the recommendation.

## 3. THE CASIF MODEL

In this section, we introduce the proposed context-aware short-term interest first model for session-based recommendation (CASIF). We first formulate the problem of session-based recommendation and then describe the architecture of our model in detail (As shown in Figure 1.)

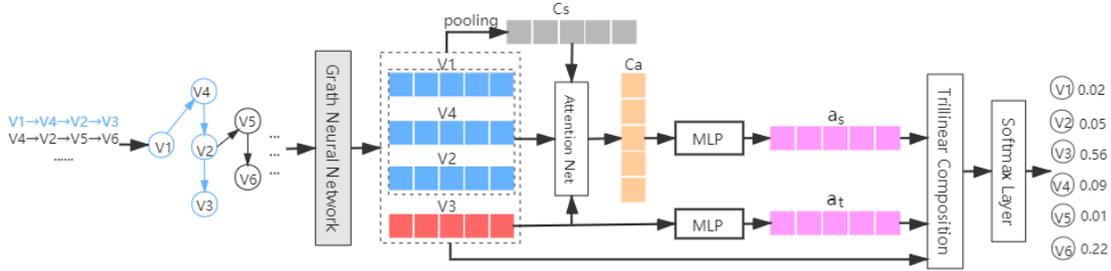

Figure 1. The general architecture of the proposed model.

## 3.1. Problem formulation

Session-based recommendation aims to predict which item a user will click next, solely based on the user's current sequential session data without access to the long-term preference profile. Here we give a formulation of this problem as below.

In session-based recommendation, let $V = \{v1, v2, ..., v_m\}$ denote the set consisting of all unique items involved in all sessions, where $m$ represents the total number of items. An anonymous session sequence $S$ can be represented by a list $S = \{s_1, s_2, .., s_n\}$ ordered by timestamps, where $s_t \in V$ represents a clicked item of the user at time step $t$. Finally, the session-based recommendation aims to predict the next possible click (i.e., $s_{t+1}$) for a given prefix of the action sequence truncated at time t, $S = \{s_1, s_2, ..., s_{t-1}, s_t\}$. Specifically, our model returns a score list $\hat{y} = \{\hat{y}_1, \hat{y}_2, \dots, \hat{y}_m\}$, where $\hat{y}$ represents the predicted scores respect to the item set $v_i$. Usually, top-k items will be chosen as recommendation items.

## 3.2. Structuring dynamic graph

The first part of the graph neural network module is to construct the meaningful graph from all the sessions. We embed every item $v \in V$ into a unified embedding space, which is represented as $s$. Given a session $S = \{s_1, s_2, \dots, s_n\}$, we treat each item $s_i$ as a node and $(s_{i-1}, s_i)$ as an edge which represents a user clicks item $s_i$ after $s_{i-1}$ in the session $S$. Therefore, every session sequence can be modeled as a directed graph. For example, considering a session $S = \{s_1, s_2, s_3, s_4, s_5\}$, the corresponding incoming matrice $\mathbf{M}^I \in \mathbb{R}^{5 \times 5}$ and outgoing matrice $\mathbf{M}^O \in \mathbb{R}^{5 \times 5}$ are shown in Figure2. Due to some items that may appear in the sequence repeatedly, we normalized weighted of all edges, which is calculated as the occurrence of the edge divided by the outdegree of that edge's start node. Note that our model can support various strategies of constructing a session graph and generate the corresponding connection matrices. Then we can apply the two weighted connection matrices with graph neural network to capture the local information of the session sequence.

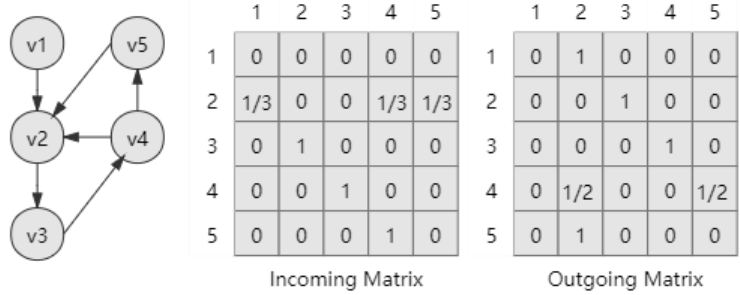

Figure 2. An example of a session graph structure and the connection matrices $\mathbf{M}^I$ and $\mathbf{M}^O$

The node vector $\mathbf{h} \in \mathbb{R}^d$ indicates the latent vector of the item $s$ learned via graph neural networks, where $d$ is the dimensionality. The process of obtaining latent feature vectors of nodes as follows:

$$\mathbf{m}_i = Concat\begin{pmatrix} \mathbf{M}_i^I([s_1, \dots, s_n] \mathbf{W}^I + \mathbf{b}^I) + \mathbf{b}_i, \\ \mathbf{M}_i^O([s_1, \dots, s_n] \mathbf{W}^O + \mathbf{b}^O) + \mathbf{b}_o \end{pmatrix}, \quad (1)$$

Where $\mathbf{M}_i^I \in \mathbb{R}^{1 \times n}$ and $\mathbf{M}_i^O \in \mathbb{R}^{1 \times n}$ represent the ith row of blocks in incoming matrices $\mathbf{M}^I \in \mathbb{R}^{n \times n}$ and outgoing matrices $\mathbf{M}^O \in \mathbb{R}^{n \times n}$ respectively corresponding to node $s_i$. $\mathbf{m}_i$ extracts the local contextual information of neighborhoods for node $s_i$. $\mathbf{W}^I, \mathbf{W}^O \in \mathbb{R}^{d \times d}$ are the parameter matrices. $\mathbf{b}^I, \mathbf{b}^O \in \mathbb{R}^d$ are the bias vectors. Eq.(1) is used for information propagation between different nodes. Next, $\mathbf{m}_i$ and $s_i$ as input of graph neural network.

$$z_i = \sigma(W_z m_i + P_z s_i), \quad (2)$$

$$r_i = \sigma(W_r m_i + P_r s_i), \quad (3)$$

$$\tilde{h}_i = \tanh(W_o m_i + P_o(r_i \odot s_i)), \quad (4)$$

$$h_i = (1 - z_i) \odot s_i + z_i \odot \tilde{h}_i, \quad (5)$$

Where $W_z, W_r, W_o \in \mathbb{R}^{2d \times d}$, $P_z, P_r, P_o \in \mathbb{R}^{d \times d}$, are learnable parameter matrices. $\sigma(\cdot)$ represents the logistic sigmoid function and $\odot$ is element-wise multiplication operator. $z_i$ and $r_i$ are the reset and update gates respectively, which determines whether some information is retained or forgotten. Finally, output $h_i \in \mathbb{R}^d$ of the GNN layer is the latent future vector corresponding to $s_i$.

Note that for the construction of dynamic graph, in practice, the graph structure containing different information can be built according to the actual situation, such as the type of item, price and other ancillary information.

### 3.3. Short-term Interest First Module

After obtaining latent future vectors from the graph neural network, we input them into short-term interest first module. As can be seen in Figure 1. the pooling operation is used to generate the session feature $m_s$, which represents the main interest of the current session. Due to the mean average of all items feature in the whole session represents a central feature of the session, there we choose the mean pooling, the specific formula is as Eq.(6).

$$c_s = \frac{1}{n} \sum_{i=1}^{n} h_i, \quad (6)$$

After capturing the user's general interest $m_s$, we use an attention network layer to obtain the attention coefficient based on short-term interest first. Attention coefficient $S$ are computed as follows:

$$\alpha_i = W_0 \, \sigma(W_1 h_i + W_2 h_n + W_3 c_s + b_a) \quad (7)$$

Where $h_i \in \mathbb{R}^d$ denotes the ith latent vector corresponding to ith item $s_i$, $h_n$ denotes the latent vector of the last click item $s_n$, $W_0 \in \mathbb{R}^{1 \times d}$ is a weighting vector, $W_1, W_2, W_3 \in \mathbb{R}^{d \times d}$ are weighting matrices, $b_a \in \mathbb{R}^d$ is a bias vector, and $\sigma(\cdot)$ denotes the sigmoid function. $\alpha_i$ represents the attention coefficient of latent vector $h_i$ within the current session prefix $S = \{h_1, h_2, ..., h_n\}$. From Eq.(7) one can see that the attention coefficients are calculated based on the latent vector $h_i$, the last-click $h_n$ and session representation $c_s$, thus, it can capture the correlations between the target item and the long/short term memory of the user's interest. Note that in Equation 7, the short-term interest is represented the last click item, which is explicitly considered, and this is why the proposed module is called the short-term interest first module.

So after the previous calculation, we can obtain the attention coefficients $\alpha = (\alpha_1, \alpha_2, ..., \alpha_n)$ with corresponding to the current session $S = \{\mathbf{h}_1, \mathbf{h}_2, ..., \mathbf{h}_n\}$, the attention-based user's global interest $\mathbf{c}_a$ can be calculated as follows:

$$\mathbf{c}_a = \sum_{i=1}^{n} \alpha_i \mathbf{h}_i, \quad (8)$$

### 3.4. MLP layer

Next, we choose the latent feature vector of the last click item $\mathbf{h}_n$ as current interest. The general interest $\mathbf{c}_a$ and current interest $\mathbf{h}_n$ are processed with two MLP networks for feature abstraction. The two MLP has the same structure except for different parameters. The specific definition is as follows:

$$a_s = \tanh(\mathbf{w}_s \mathbf{c}_a + \mathbf{b}_s), \quad (9)$$

$$a_t = \tanh(\mathbf{w}_t \mathbf{c}_a + \mathbf{b}_t), \quad (10)$$

Where $a_s, a_t \in \mathbb{R}^d$ denotes the final state of global interest and current interest respectively, $\mathbf{w}_s, \mathbf{w}_t \in \mathbb{R}^{d \times d}$ are learnable weighting matrix, and $\mathbf{b}_s, \mathbf{b}_t \in \mathbb{R}^d$ are the bias vector. The tanh is non-linear activation function.

### 3.5. Model training and making recommendation

Finally, for a given candidate item $s_i \in V$, the score function is defined as:

$$\hat{z}_i = s_i^T (a_s \odot a_t), \quad (11)$$

And then we apply a softmax function to get the output vector of the model $\hat{y}$:

$$\hat{y} = softmax(\hat{z}), \quad (12)$$

Where $\hat{z} \in \mathbb{R}^{|V|}$ denotes the recommendation scores with respect to the item set $V$, and $\hat{y}$ represents a probability distribution over the items $s_i \in V$, each element $\hat{y}_i \in \hat{y}$ denotes the probability of the event that item $s_i$ is going to appear as the next-click in this session.

For each session prefix $S = \{s_1, s_2, s_3, s_4, s_5\}$, the loss function is defined as the cross-entropy of the prediction and the ground truth:

$$\mathcal{L}(\hat{y}) = -\sum_{i=1}^{|V|} y_i \log \hat{y}_i + (1 - y_i) \log(1 - \hat{y}_i), \quad (13)$$

Where $y$ denotes a one-hot vector of the ground truth item. For example, if $s_{n+1}$ denotes the ith element $v_i$ in item dictionary $V$, then $y_k = 1$, if $i == k$, and $y_k = 0$.

## 4. EXPERIMENTAL RESULTS AND ANALYSIS

### 4.1. Model training and making recommendation

#### 4.1.1. Datasets

Our experiment constructs on two real-world datasets: Yoochoose from RecSys Challenge 2015 and Diginetica from CIKM Cup 2016. To facilitate comparison, we preprocess the two datasets in the same way as [6,8]. First, we process items in all sessions in chronological order. Second, we filter out those sessions that have only one item and that have items appearing less than 5 times. Third, we generate the sequences and the corresponding lables by splitting the input sequence. Specifically, we set the sessions for the next few days as test sets for Yoochoose and the sessions for the next few weeks as test sets for Diginetiva. we also use the most recent fractions 1/64 and 1/4 of the training sequences of Yoochoose. The statistics of datasets are presented in Table 1.

Table 1. Statistics of datasets used in the experiments

| Statistics | Yoochoose1/64 | Yoochoose1/4 | Diginetica |
|---|---|---|---|
| all the clicks | 557,248 | 8,326,407 | 982,961 |
| train sessions | 368,859 | 5,917,745 | 719,470 |
| test sessions | 55,898 | 55,898 | 60,858 |
| all the items | 16,766 | 29,618 | 43,097 |
| average length | 6.16 | 5.71 | 5.12 |

#### 4.1.2. Baselines

We compare the proposed models with four traditional methods (POP, IKNN, BPR-MF, FPMC) and three recent deep learning models (GRU4REC, NARM, STAMP, SR-GNN).

- Pop is a simple baseline that recommends top rank items based on popularity in training data.
- Item-KNN [30] recommends the item similar to the items that have been clicked in the current session, where the cosine similarity is used.
- BPR-MF [28] is a learning-to-rank method. It is the most advanced non-sequential recommended method, which utilizes pair-sort loss optimization matrix decomposition.
- FPMC [2] combines the Markov chain model and matrix factorization for the next-basket recommendation.
- GRU4REC [4] is an RNN-based deep learning model for session-based recommendation. It utilizes a session-parallel mini-batch training process and applies a ranking-based loss function for training.
- NARM [6] adopts recurrent neural network as its basic component and utilizes an attention mechanism to extract users' main purpose.
- STAMP [8] is a novel short-term memory priority model. The attention mechanism is used to capture the user's general interest, and finally the click item represents the user's current interest.

- SR-GNN [27] models session sequences as graph structure data and uses graph neural networks to obtain item latent vectors, which are input to a traditional attentive neural network for learning session representation.

### 4.1.3. Performance metrics

We use the following performance metrics to compared these algorithms, which have been widely used in session-based recommendation systems.

$Recall@k$: Be widely used as a measure of predictive accuracy in all kinds of recommendation systems. It represents the proportion of correctly recommended items amongst the top-k items.

$$Recall@k = \frac{n_{hit}}{N}, \quad (14)$$

Where $N$ is the number of test sessions in the testing set, $n_{hit}$ denotes the number of sessions which have hit items among top-K ranking list.

$MRR@k$: MRR (Mean Reciprocal Rank) is the average of reciprocal ranks of desired items. The reciprocal rank is set to zero if the rank is larger than K.

$$MRR@k = \frac{1}{N}\sum_{i=1}^{N}\frac{1}{rank_i}, \quad (15)$$

### 4.1.4. Parameter settings

Following previous methods [6, 8, 28] the dimension of embedding D is set to 100 for both datasets. All parameters are initialized using a Gaussian distribution with a mean of 0 and a standard deviation of 0.1. The proposed CASIF model uses Adam optimizer to optimize these parameters and the batch size is set as 128. For the Yoochoose, the learning rate is set to 0.001. For the Diginetica, the learning rate is 0.003, and all learning rates will decay by 0.1 after every 3 epochs. Besides, the L2 penalty is 10e-5. Lastly, our model is implemented with Pytorch on GeForce GTX 1660Ti GPU and all the experimental results are the average value of five times testes.

## 4.2. Experiment results

### 4.2.1. Comparison with Baseline Methods

To state the performance of our CASIF model for session-based, all experimental results were evaluated by Recall@20 and MRR@20. As shown in Table 2, the best results have been highlighted in boldface. Results analysis is mainly divided into three parts, the first is the traditional method, the second is the deep learning related models, and the last is our model.

For traditional models, POP, as the simplest algorithm, has the worst recommendation performance. By analyzing the users individually and optimizing the paired ranking loss function, the performance of BPR-MF is better than that of POP. This suggests the importance of personalization in the recommendation task. Although FPMC integrates Markov chain and matrix decomposition, the overall result is not as good as Item-KNN. Please note that Item-KNN only uses the similarity between items, and does not consider the sequence information. This shows that the assumption of independence of continuous terms that the traditional MC method relies on is not realistic.

Obviously, the performance of all neural network models is better than that of traditional methods. This also verifies the powerful role of deep learning in this field, because they can extract some deep-seated and representative potential features from the temporal relationship of items in historical sessions. GRU4REC uses the recursive structure GRU as a special form of RNN to capture the general preferences of users. It improves the performance of the session-based recommendation greatly. But NARM and STAMP are better than it, which shows the effectiveness of attention mechanism and short-term behavior in predicting the next project problem.

Our model just uses a deep learning method and attention mechanism. Therefore, compared with the baseline model, our method achieve the best results. First, we use graph-structured data to input into the neural network, which captures the local dependence of the session. Secondly, in the attention layer, we first consider the short-term behavior and then integrate global behavior, which captures more accurate context information. Finally, more potential features are extracted by MLP. Our model is especially suitable for recommendation tasks with a large amount of data.

Table 2. Performance comparison for different methods over the three datasets.

| Dataset | Yooochoose 1/64 | | Yooochoose 1/4 | | Diginetica | |
|---|---|---|---|---|---|---|
| Measure | Recall@20 | MRR@20 | Recall@20 | MRR@20 | Recall@20 | MRR@20 |
| POP | 6.71 | 1.65 | 1.33 | 0.30 | 0.91 | 0.23 |
| BPR-MF | 31.31 | 12.08 | 3.40 | 1.57 | 15.19 | 8.63 |
| IKNN | 51.60 | 21.81 | 52.31 | 21.70 | 28.35 | 9.45 |
| FPMC | 45.62 | 15.01 | -- | -- | 31.55 | 8.92 |
| GRU4REC | 60.64 | 22.89 | 59.53 | 22.60 | 43.82 | 15.46 |
| NARM | 68.32 | 28.63 | 69.73 | 29.23 | 62.58 | 27.35 |
| STAMP | 68.74 | 29.67 | 70.44 | 30.00 | 62.03 | 27.38 |
| SR-GNN | 70.57 | 30.94 | 71.36 | 31.89 | 63.03 | 27.42 |
| CASIF | **70.70** | **31.21** | **72.01** | **32.11** | **63.59** | **28.33** |

### 4.2.2. Further comparison with excellent baselines

In order to further study the performance of our proposed CASIF and the state-of-the-art methods NARM, STAMP and SR-GNN in the real application environment, where only a few items can be recommended to users at once. Therefore, we evaluate our model in terms of Recall@10, MRR@10, Recall@5, and MRR@5 to further measure recommendation accuracy. The results on two datasets Yoochoose1/64 and Diginetica are summarized in table 3. As we can see, Our model is comparable to SR-GNN in terms of smaller data sets Yoochoose1/64 and Recall@5, which mainly because the data set is limited and the precision is not easy to improve on the strict performance metrics. However, overall, our model is comparable performs well on this case and shows obvious advantages, especially on MRR@10, which indicates that our model is more accurate in ranking candidate items and demonstrates the effectiveness of taking into account both the context information and shor-term intersts first for session-based recommendation.

Table 3. Recommendation performance of STAMP, SR-GNN and our
CASIF on Recall@k, MRR@k, where k=10, 5

| Dataset | Measure | NARM | STAMP | SR-GNN | CASIF | Improve |
|---|---|---|---|---|---|---|
| Yoochoose1/64 | Recall@10(%) | 57.50 | 58.07 | 60.01 | 61.21 | +1.99% |
| | MRR@10(%) | 27.97 | 28.92 | 29.56 | 30.58 | +3.45% |
| | Recall@5(%) | 44.34 | 45.69 | 47.11 | 46.88 | -0.49% |
| | MRR@5(%) | 26.21 | 27.26 | 28.18 | 28.29 | +0.03% |
| Diginetica | Recall@10(%) | 51.91 | 52.07 | 52.70 | 53.18 | +0.91% |
| | MRR@10(%) | 26.53 | 26.90 | 27.12 | 27.86 | +2.72% |
| | Recall@5(%) | 40.67 | 41.04 | 41.45 | 41.52 | +0.16% |
| | MRR@5(%) | 25.02 | 25.21 | 25.42 | 26.04 | +2.43% |

**4.2.3. Effectiveness of short-term interest first module**

To verify the effectiveness of short-term interest first module, We design a variant of our model, that is, after GNN, as shown in Eq. (16, 17, 18), we directly obtain the attention coefficient in a simple attention, and then sum weighted by the corresponding latent vectors. We mark this variant as CASIF-S. From Fig. 3. , we can see the experimental results, whether on the evaluation metrics Recall@20 or MRR@20, the performance of CASIF is higher than CASIF-S. the experience verifies the importance of the short-term interest first module in our proposed model, and it also demonstrates that it is not enough to only consider GNN, because GNN captures only the local dependencies of the session and does not obtain the main intent of the session from global. Especially, the pooling layer in our model can capture the main interst of session. And only consider attention, easy to deviate from the overall.

$$\alpha_i = \mathbf{W}_0 \, \sigma(\mathbf{W}_1 \mathbf{h}_i + \mathbf{b}_1), \quad (16)$$

$$\mathbf{h}_a = \sum_{i=1}^{n} \alpha_i \mathbf{h}_i, \quad (17)$$

$$\hat{z}_i = s_i^T \mathbf{h}_a, \quad (18)$$

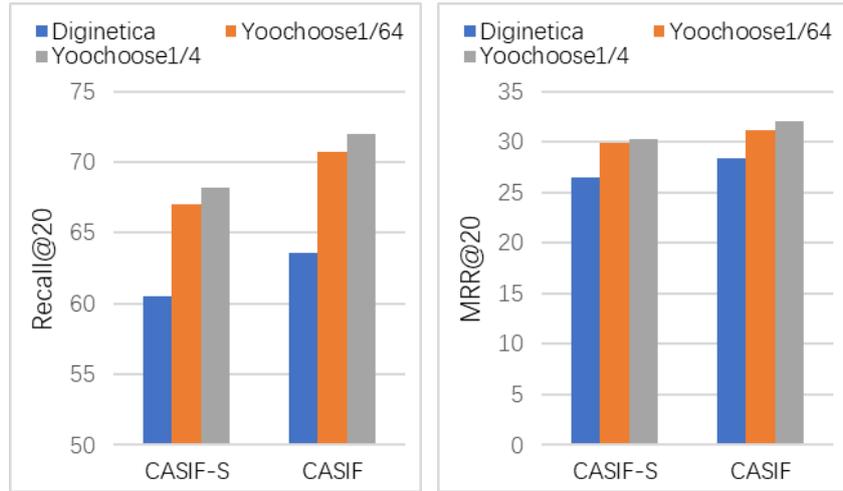

Figure 3. Recommendation performance of CASIF-S and CASIF in terms of Recall@20 (left) and MRR@20 (right) on three real-world datasets.

### 4.2.4. Analysis on Session Sequence Length

In order to compare the performance of the model on different session length datasets. We divide Yoochoose1/64 and Diginetica into two groups respectively. One is a short session datasets, which is marked as "Short", and the session length is less than or equal to 5. The other group of datasets has each session length greater than 5, which is named as "Long". We chose representative, advanced model (STAMP, SR-GNN) as baselines. The experimental results under Recall@20 and MRR@20 are shown in table 4 and table 5 respectively. The best results have been highlighted in boldface. As we can see, our model performs excellent on both the "Long" dataset and the "Short" dataset. On the "Short" dataset, our model is comparable to SR-GNN, which thanks to the power of graph neural networks. On the whole, our method is superior, it is mainly due to the fact that our model is mainly composed of GNN module and short-term interest first module. GNN can perform well on long session sequences, while short-term interest first module can perform well on short session sequences. It can be seen that our model is suitable for both recommendation tasks for long sessions sequences and recommendation tasks for short sessions sequences.

Table 4. The performance of different methods with different session lengths evaluated in terms of Recall@20

| Method | Yoochoose1/64 | | Diginetica | |
|---|---|---|---|---|
| | Short | Long | Short | Long |
| STAMP | 71.44 | 64.73 | 47.26 | 40.39 |
| SR-GNN | 70.69 | 70.70 | 50.49 | 21.27 |
| CASIF-S | 69.52 | 70.68 | 50.32 | 20.89 |
| **CASIF** | **71.56** | **70.88** | **51.12** | **51.36** |

Table 4. The performance of different methods with different session lengths evaluated in terms of MRR@20

| Method | Yoochoose1/64 | | Diginetica | |
|---|---|---|---|---|
| | **Short** | **Long** | **Short** | **Long** |
| **STAMP** | 32.60 | 24.31 | 26.26 | 25.33 |
| **SR-GNN** | **31.15** | 30.93 | 27.49 | 26.27 |
| CASIF-S | 29.52 | 28.78 | 26.45 | 25.33 |
| CASIF | 31.02 | **31.28** | **28.12** | **29.83** |

## 5. CONCLUSIONS

In this paper, we propose context-aware short-term interest first model for session-based recommendation, which combines GNN with short-term attention priority. GNN is used to capture the local dependencies of the session, while the short-term interest first module fully considers the current interest and long-term interest, which can extract the main intention of the session. We make full use of the complementarity of the two, and a large number of experimental results on two data sets demonstrate the superiority of our proposed model. However, the running time of this model is not optimized compared with previous methods. In the following work, we will continue to try to improve the structure of the model to make the model less complex and more efficient.

## ACKNOWLEDGEMENTS

I would like to thank the tutor for his guidance on my research direction and the conference organizers and reviewers!

AUTHORS

**Haomei Duan** is currently a master student in the College of Computer Science and Technology, Heilongjiang University, Harbin, China. She received the B.S. degree in College of Computer Science and Technology from Shandong University of Technology. Her research focuses on recommendation algorithms and systems.

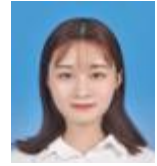

**Jinghua Zhu** received the B.S. degree in computer software and the M.S. degree in computer science in 1999 and 2002, respectively, and the Ph.D. degree in computer science from Harbin Institute of Technology in 2009. She has been a Professor with the School of Computer Science and Technology, Heilongjiang University, China, since 2016. She has published many high quality conference and journal research papers. Her research interests include social networks, data mining, uncertain databases, and wireless sensor networks.

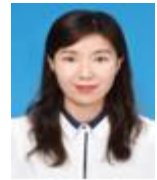